\documentclass[prl,english,aps,superscriptaddress,twocolumn,floatfix]{revtex4-1}
\usepackage{epsf}
\usepackage{graphicx}
\usepackage{sidecap}
\usepackage{amsmath}
\usepackage{amssymb}
\usepackage{bm}

\usepackage{color}

\def \beq {\begin{equation}}
\def \eeq {\end{equation}}
\begin{document}

\title{{Theory of Out-of-Equilibrium Ultrafast Relaxation Dynamics in Metals}}

\author{Pablo Maldonado}
\affiliation{Department of Physics and Astronomy, Uppsala University, P.\,O.\ Box 516, S-75120 Uppsala, Sweden}
\author{Karel Carva}
\affiliation{Charles University, Faculty of Mathematics and Physics,
Department of Condensed Matter Physics, Ke Karlovu 5, CZ-12116 Prague 2, Czech Republic}
\affiliation{Department of Physics and Astronomy, Uppsala University, P.\,O.\ Box 516, S-75120 Uppsala, Sweden}
\author{Martina Flammer}
\affiliation{Department of Physics and Astronomy, Uppsala University, P.\,O.\ Box 516, S-75120 Uppsala, Sweden}
\author{Peter M.\ Oppeneer}
\affiliation{Department of Physics and Astronomy, Uppsala University, P.\,O.\ Box 516, S-75120 Uppsala, Sweden}

\begin{abstract}
\noindent
Ultrafast laser excitation of a metal causes correlated, highly nonequilibrium dynamics of electronic and ionic degrees of freedom, which are however only poorly captured by the widely-used two-temperature model.
Here we develop an out-of-equilibrium theory that 
captures the full dynamic evolution of the electronic and phononic populations and provides a microscopic description of the transfer of energy delivered optically into electrons to the lattice. All essential nonequilibrium energy processes, such as electron-phonon and phonon-phonon interactions are taken into account.
Moreover, as all required quantities are obtained from first-principles calculations, the model gives an exact description of the relaxation dynamics without the need for fitted parameters.
We apply the model to FePt and show that the detailed relaxation is out-of-equilibrium for picoseconds.
\end{abstract}
\date{\today}
\maketitle

Excitation of a material with an intensive, ultrashort optical pulse brings the material's electrons into a strongly out-of-equilibrium state which is immediately followed by intense, correlated dynamics between the electrons and other degrees of freedom in the material, such as the lattice or spin systems \cite{Fujimoto1984,Beaurepaire1996,Koopmans2010,Liao2016}.
The ability to measure the ultrafast relaxation dynamics of the involved subsystems using pump-probe techniques has led to the discovery of many unexpected phenomena, such as ultrafast demagnetization \cite{Beaurepaire1996}, change of magnetic anisotropy \cite{Hansteen2005}, or coherent generation of magnetic precession \cite{Ju1999,VanKampen2002}.
More recently, ultrafast generation of lattice strain waves \cite{Kim2012,Kim2015,Henighan2016}, coherent control of atomic and molecular dynamics \cite{Rabitz2008}, 
coherent phonon  generation  \cite{Harmand2013,Lindenberg2017}, and laser-induced superconductivity at high temperature \cite{Mitrano2016} have been reported. In spite of the importance of the material's nonequilibrium state in these laser-induced phenomena, it is surprising that most theoretical descriptions 
of the ensuing out-of-equilibrium dynamics are based on the widely-used two-temperature model (2TM) \cite{Kaganov1957,Anisimov1974}, 
which assumes that the electronic and phononic subsystems are separately in thermodynamic equilibrium \cite{Allen1987}.

Research on nonequilibrium states of matter has emerged recently as an important area in condensed matter physics (see, e.g., \cite{Stojchevska2014,Aoki2014}), and consequently several improved models have been developed which incorporate aspects of out-of-equilibrium electronic dynamics \cite{Carpene2006,Mueller2013,Carva2013,Baranov2014}.  However, these still lack a complete out-of-equilibrium description of the full system and its time evolution, and contain parameters that are either fitted experimentally or chosen from a macroscopic system at equilibrium. 
Additionally, recent investigations \cite{Henighan2016,Waldecker2016,Waldecker2017} have emphasized that the assumption of thermal phonons 
could lead to marked disagreement with experimental observations.

Here we propose a general theory to describe the ultrafast dynamics triggered by ultrashort laser pulses in metals. Contrarily to the two-temperature model, our model employs an out-of-equilibrium description of the electronic and phononic populations and provides the full dynamic description of the nonequilibrium relaxation processes. The decisive quantities that govern the dynamical relaxation 
are the phonon mode dependent electron-phonon coupling constants and the mode dependent phonon-phonon scattering terms.
It is important to stress that both quantities have an explicit dependence on the electronic temperature, the phonon momentum and branch, 
and, moreover, that they are fully derived from microscopic \textit{ab initio} theory.
Thus, as the theory only uses quantities obtained from \textit{ab initio} calculations, it provides a \textit{parameter free} description of the relaxation dynamics and could hence become of great value for modeling and predicting nonequilibrium dynamics following ultrafast laser excitation.
As an example, the model is used to describe the ultrafast dynamics in FePt after femtosecond laser irradiation, illustrating as well the limitations of the 2TM. 


\begin{figure}[th]
\vspace*{-0.2cm}
\includegraphics[width=0.9\linewidth]{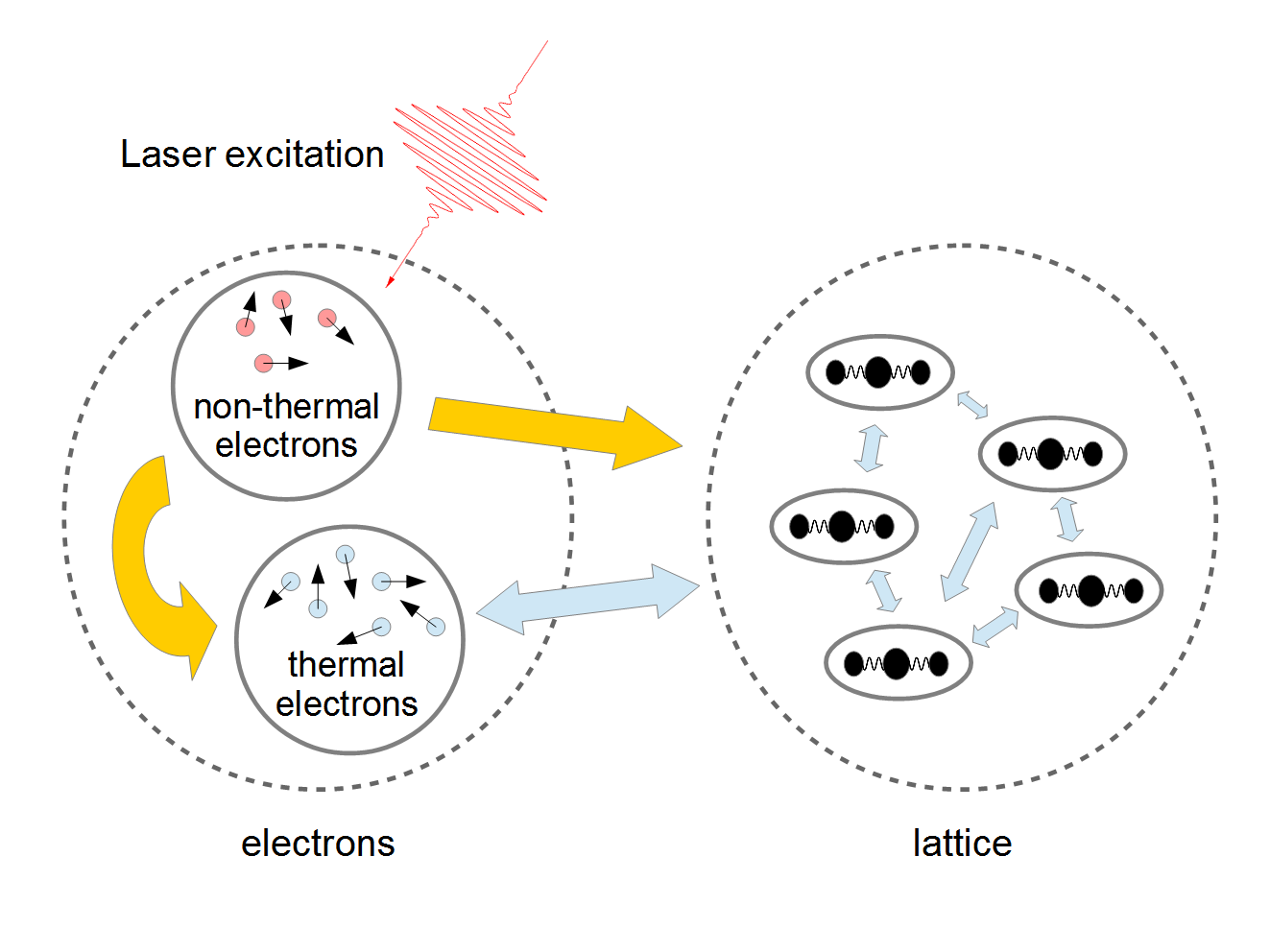}
\caption{Scheme of treating out-of-equilibrium dynamics of electrons and lattice. The electronic subsystem is excited via femtosecond laser pulses. The major part of the electronic system is considered in local thermal equilibrium whereas a small part that absorbs most of the laser energy is in a nonthermal state. The electronic subsystems relax energy by interaction with different phononic subsystems, which also exchange energy among themselves (depicted by arrows).}
\label{Figschem}
\end{figure}

To describe the nonequilibrium time evolution of the electronic and phononic degrees of freedom of the laser-excited material,
we divide the out-of-equilibrium metallic system in different, independent subsystems that interact with one another, schematically shown in Fig.\ \ref{Figschem}.
Specifically, we describe the lattice by dividing it into $N$ independent phonon subsystems, each of them corresponding to a specific branch $\nu$ and momentum $\bm{q}$. They interact with one another through phonon-phonon scattering and with the electrons via electron-phonon scattering. These interactions are phonon momentum and branch dependent. Therefore these phonon subsystem populations $n_{\nu \bm{q}}$ evolve separately during the nonequilibrium dynamics 
and we can define a ``lattice temperature" $T_{\ell}^Q$ (with $Q \equiv \nu\bm{q}$) for each of them
(the meaning of this definition is discussed further below). 
The impulsive laser excitation brings a part of the electrons into a nonthermal state (Fig.\ \ref{Figschem}). 
For the electrons we follow the description of Carpene \cite{Carpene2006}, in which the electronic system is divided in a thermal bath that contains the majority of thermal electrons with temperature $T_e$ and in a laser-excited nonthermal electron distribution, which relaxes driving energy into the thermal distribution through the electron-electron and electron-phonon interactions. The latter interaction conducts the energy from the laser-excited electrons to the different lattice subsystems, bringing them to different temperatures, i.e., into a nonequilibrium state. The phonon-phonon scattering causes the transferred energy to be shared between the phonon subsystems, guiding them toward a common lattice temperature, and therefore to a thermal equilibrium of the lattice. Hence, the rates of electron-phonon and phonon-phonon scattering 
are the quantities that determine the system's temporal evolution to equilibrium.




To achieve a theoretical formulation we make use of the conservation of total energy and classical kinetic theory. 
The total lattice energy is given by $E_{\ell} =\sum_Q\hbar\omega_Q n_Q$ and the total electronic energy by $E_e=2\sum_k\epsilon_k f_k$, where  $\hbar \omega_Q$ is the phonon energy, $\epsilon_{k}$ the electron Bloch energy \cite{note1} (with $k \equiv n\bm{k}$), and $n_Q$ and $f_k$ are phonon and electron populations, respectively.
The latter are in local equilibrium given by the Fermi-Dirac and Bose-Einstein distributions
\begin{equation}
f_k \!= \!{\Big[{e^{\frac{\epsilon_{k}-\epsilon_{_F}}{k_{B} T_e (t)}}+1} \Big]^{-1}}, ~{\rm and}~~
n_Q \!=\!{\Big[{e^{\frac{\hbar\omega_{Q}}{k_{B} T_{\ell}^Q (t)}}-1} \Big]^{-1}} .
\end{equation}
We can define the rates of energy exchange as
\begin{eqnarray}
\frac{\partial E_{\ell}}{\partial t}&=&\sum_Q \hbar\omega_Q \dot{n}_Q \big|_{e-ph}^{scatt.}+\sum_Q \hbar\omega_Q \dot{n}_Q \big|_{ph-ph}^{scatt.},\label{rate-E1}\\
\frac{\partial E_e}{\partial t}&=& 2\sum_k \epsilon_k \dot{f}_k \big|_{e-ph}^{scatt.} = -\sum_Q \hbar\omega_Q \dot{n}_Q \big|_{e-ph}^{scatt.},
\label{rate-E2}
\end{eqnarray}
where the dot stands for the time derivative and the subscripts denote the different scattering processes that change the distribution. 
The equivalence in Eq.\ (\ref{rate-E2}) stems from the conservation of total energy.
The time derivatives of the distribution functions due to different scattering terms can be derived from the classical kinetic theory by using the well-known Fermi's golden rule of scattering theory. By doing so, we obtain an extended version of the Bloch-Boltzmann-Peierls equations (see \cite{Ziman-1960}, and Supplemental Material (SM) \cite{SM}),
\begin{widetext}
\begin{eqnarray}
\dot{f}_k\big|_{e-ph}^{scatt.} &=& 
 -\frac{2\pi}{\hbar}\sum_Q |M_{kk^{\prime}}|^2 \displaystyle{ \left\{ f_k(1-f_{k^{\prime}})\left[ (n_Q+1)\delta (\epsilon_k-\epsilon_{k^{\prime}}-\hbar\omega_Q)+n_Q\delta (\epsilon_k-\epsilon_{k^{\prime}}+\hbar\omega_Q)\right] \right.} \nonumber \\
& & \displaystyle{\left. -(1-f_k)f_{k^{\prime}} \left[ (n_Q+1)\delta (\epsilon_k-\epsilon_{k^{\prime}}+\hbar\omega_Q)+n_Q\delta (\epsilon_k-\epsilon_{k^{\prime}}-\hbar\omega_Q)\right]  \right\} } , \label{BBP1} \\
\dot{n}_Q\big|_{e-ph}^{scatt.}&=&-\frac{4\pi}{\hbar}\sum_Q |M_{kk^{\prime}}|^2 f_k(1-f_{k^{\prime}})\left[ n_Q \delta (\epsilon_k-\epsilon_{k^{\prime}}+\hbar\omega_Q)-(n_Q+1)\delta (\epsilon_k-\epsilon_{k^{\prime}}-\hbar\omega_Q)\right] , \label{BBP2}\\
\dot{n}_Q\big|_{ph-ph}^{scatt.} & = &\frac{2\pi}{\hbar}\sum_{k^{\prime}k^{\prime\prime}}|\Phi_{-Qk^{\prime}k^{\prime\prime}}|^2 \left\{ (n_Q+1)(n_{k^{\prime}}+1)n_{k^{\prime\prime}}\delta (\omega_Q+\omega_{k^{\prime}}-\omega_{k^{\prime\prime}})+ \right.  \nonumber\\
&& (n_Q+1)(n_{k^{\prime\prime}}+1)n_{k^{\prime}}\delta (\omega_Q+\omega_{k^{\prime\prime}}-\omega_{k^{\prime}}) -n_Q n_{k^{\prime}}(n_{k^{\prime\prime}}+1) \delta (\omega_Q+\omega_{k^{\prime}}-\omega_{k^{\prime\prime}}) \nonumber\\
&& -n_Qn_{k^{\prime\prime}}(n_{k^{\prime}}+1) \delta (\omega_Q+\omega_{k^{\prime\prime}}-\omega_{k^{\prime}})+(n_Q+1)n_{k^{\prime}}n_{k^{\prime\prime}} \delta (\omega_Q-\omega_{k^{\prime\prime}}-\omega_{k^{\prime}})\nonumber\\
&& \left. -n_Q(n_{k^{\prime}}+1)(n_{k^{\prime\prime}}+1) \delta (\omega_Q-\omega_{k^{\prime\prime}}-\omega_{k^{\prime}}) \right\}.
\label{BBP3}
\end{eqnarray}
\end{widetext}
Here $M_{kk^{\prime}}$ and $\Phi_{-Qk^{\prime}k^{\prime\prime}}$ are the electron-phonon and phonon-phonon matrix elements, respectively. 

Equations (\ref{rate-E1}) and (\ref{rate-E2}) describe the energy flow between electron and phonon subsystems under the assumption that the diffusion term can be neglected, which is a valid assumption on the short time scale of the typical out-of-equilibrium process.
The laser driving field induces the nonequilibrium electronic distribution and enters in the rate equations as source term. 
Substituting Eqs.\ (\ref{BBP1})--(\ref{BBP3}) in (\ref{rate-E1}) and (\ref{rate-E2}), and making an expansion of the distribution functions to second order in the phonon mode and electron temperature differences (see SM \cite{SM}), we obtain a set of coupled differential equations which connect the time evolutions of the electron temperature  $T_e$ and the temperatures $T_{\ell}^Q$ of the different phonon modes,
%
%
\begin{widetext}
\begin{eqnarray}
\! \! C_Q \frac{\partial T_{\ell}^Q}{\partial t}&=&-G_Q(T_{\ell}^Q-T_e) \! \left[1+J(\omega_Q,T_{\ell}^Q)(T_{\ell}^Q-T_e)\right]-\frac{1}{9}\sum_{k^{\prime}}C_Q\Gamma_{Qk^{\prime}}\big( T_{\ell}^Q-T_{\ell}^{k^{\prime}}\big)+\frac{1}{N_Q}\frac{\partial U_{e-ph}}{\partial t} ,  \label{rate-T1} \\
\! \! C_e\frac{\partial T_e}{\partial t}&=&\sum_Q  G_Q(T_{\ell}^Q-T_e) \! \left[1+J(\omega_Q,T_{\ell}^Q)(T_{\ell}^Q-T_e)\right]+\frac{\partial U_{e-e}}{\partial t}, 
~~~~Q=Q_1, \ldots, Q_N.
\label{rate-T2}
\end{eqnarray}
\end{widetext}
%
$C_e$ and $C_Q$ are the temperature-dependent electronic and phonon mode dependent specific heats, respectively, and $G_Q$ and $\Gamma_{Qk^{\prime}}$ are the mode-dependent electron-phonon coupling constants and the mode-dependent phonon linewidths which are caused by the phonon-phonon interactions.  $\frac{\partial U_{e-e}}{\partial t}$ and $\frac{\partial U_{e-ph}}{\partial t}$ describe the rate transferred from the laser-induced nonequilibrium electron distribution into the thermal electronic bath via electron-electron interaction and into the lattice through electron-phonon interaction, respectively \cite{Carpene2006}.
$J(\omega_Q,T_{\ell}^Q)$ is a function of the mode-dependent phonon frequencies and temperatures, which accounts for the second-order term in temperature differences;  its full form is given in the SM \cite{SM}.

To obtain a full solution of the out-of-equilibrium model, defined by Eqs.\ \eqref{rate-T1} and \eqref{rate-T2}, it is necessary to compute the material specific quantities, the phonon and electronic specific heats and phonon mode-dependent electron-phonon and phonon-phonon linewidths.  These can be conveniently calculated using the spin-polarized density functional theory (DFT) in the local-density approximation (LDA). Here we have employed the electronic structure code ABINIT \cite{ABINIT}. Specifically, the mode-dependent electron-phonon linewidths were computed as response function within the density functional perturbation theory 
whereas the mode-dependent phonon-phonon linewidths due to phonon-phonon interaction were determined using many-body
perturbation theory in a third-order anharmonic Hamiltonian which includes up to three-phonon scattering \cite{Togo2015}. The coupled  equations \eqref{rate-T1} and \eqref{rate-T2} are then solved numerically, with the \textit{ab initio} quantities, and without
any free fitting parameters.

We emphasize that the phonon branch and wavevector $\bm{q}$ dependent phonon temperatures  act here as an auxiliary quantity that allows us to use for each  phononic subspace a Bose-Einstein distribution with a local temperature $T_{\ell}^Q$. While this temperature might not be measurable, recent electron diffraction experiments showed that nonequilibrium phonon populations in reciprocal space can be measured \cite{chase2016}. The electron temperature $T_e$ is conversely a quantity that can be obtained from pump-probe photoemission measurements \cite{sun93,guo01,Lisowski04}. As will become evident below, the out-of-equilibrium model [Eqs.\ (\ref{rate-T1}) and (\ref{rate-T2})] 
leads to results that are markedly different from those of the 2TM. Nonetheless, it can be recognized that under simplifying assumptions the 2TM can be obtained from Eqs.\ (\ref{rate-T1}) and (\ref{rate-T2}). To this end, it is important to realize that $G_Q$ is the nonequilibrium equivalent of the electron-phonon coupling constant $G_{ep}$ that is commonly used in the 2TM. Assuming a single phonon temperature $T_{\ell}$ for all phonon modes and wavevectors and neglecting quadratic terms in the temperature difference  $T_e -T_{\ell}$, along with setting $U_{e-ph}=0$, and using $G_{ep}=\frac{1}{N_Q}\sum_QG_Q$ (valid under the assumptions of the 2TM, see SM), one recovers the common 2TM \cite{Anisimov1974}.


\begin{figure}[b!]
  \vspace*{-0.75cm}
\includegraphics[width=0.9\linewidth]{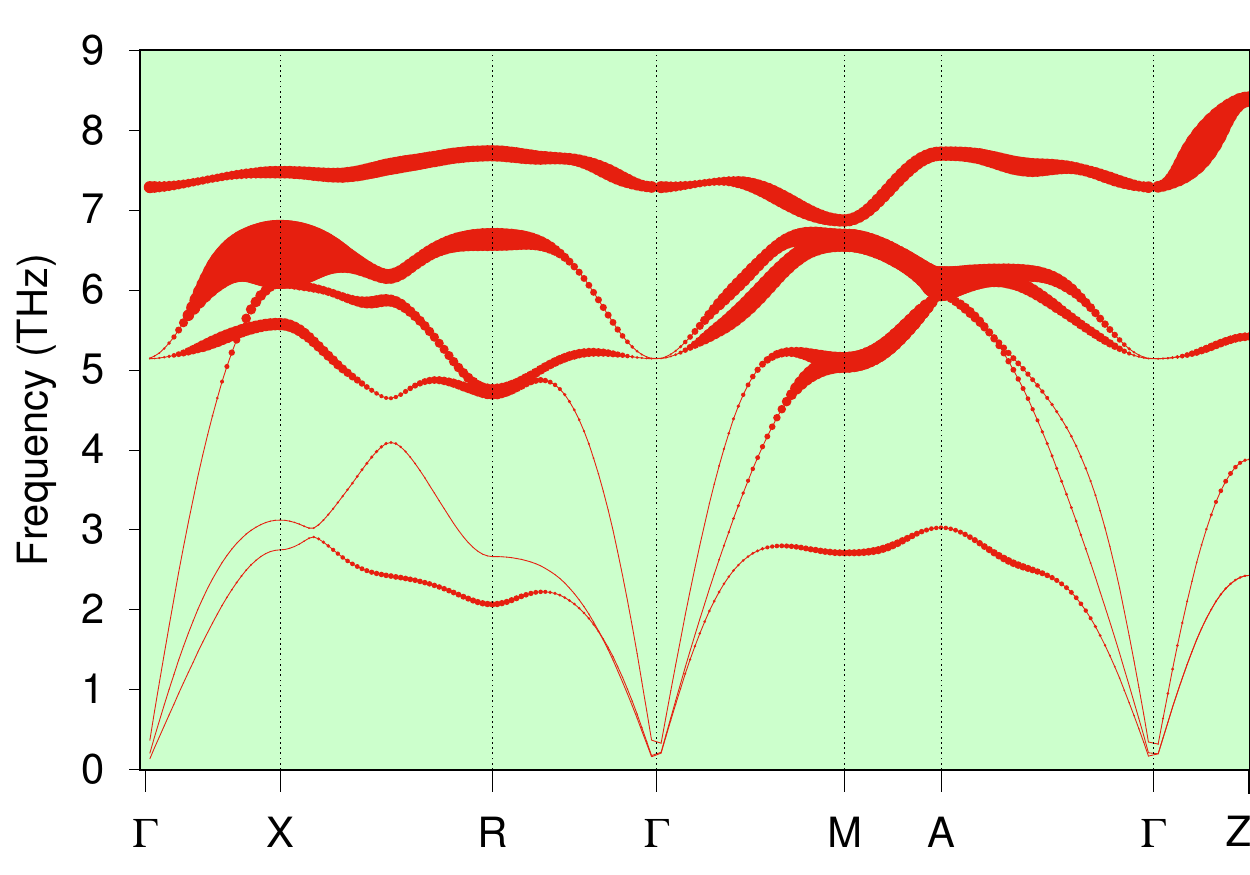}
\caption{Calculated phonon dispersions of ferromagnetic FePt along high-symmetry lines in the simple tetragonal BZ. The symbol size is proportional to the magnitude of the mode-dependent electron-phonon coupling constant $G_Q$ at 300 K. }
\label{Fig1}
\end{figure}

To recognize the importance of the mode and wavevector dependent electron-phonon coupling constant $G_Q$ we perform \textit{ab initio} calculations of this quantity for ferromagnetic FePt, which is the prime material for high-density optic and magnetic recording \cite{lambert14,john2017}. Bulk FePt orders in the L1$_0$ structure in which the (001) planes 
are alternatively  occupied  by Fe and  Pt  atoms. Our \textit{ab initio} calculation of the ground state magnetic properties of FePt is in good agreement  with previous work 
\cite{FePt}.

In Fig.\ \ref{Fig1} we show our \textit{ab initio} calculated mode-dependent electron-phonon coupling constants  $G_Q$ of FePt at 300 K together with the phonon dispersions along high-symmetry lines in the simple tetragonal Brillouin zone (BZ).
We can readily see that the electron-phonon coupling constants are larger for optical phonon modes, reaching several orders of magnitude differences between some points of the BZ (see e.g.\ optical phonons at the X point compared to acoustic phonons at the $\Gamma$ point). 
These findings demonstrate
 the limitations of considering a single $G_{ep}$ with the lattice at one local thermal equilibrium, since the range of values of $G_Q$ is several orders of magnitude. 
 On account of the different coupling strengths the laser-excited electrons will couple mainly to the optical phonon modes.
 
 \begin{figure}[t!]
\includegraphics[width=0.99\linewidth]{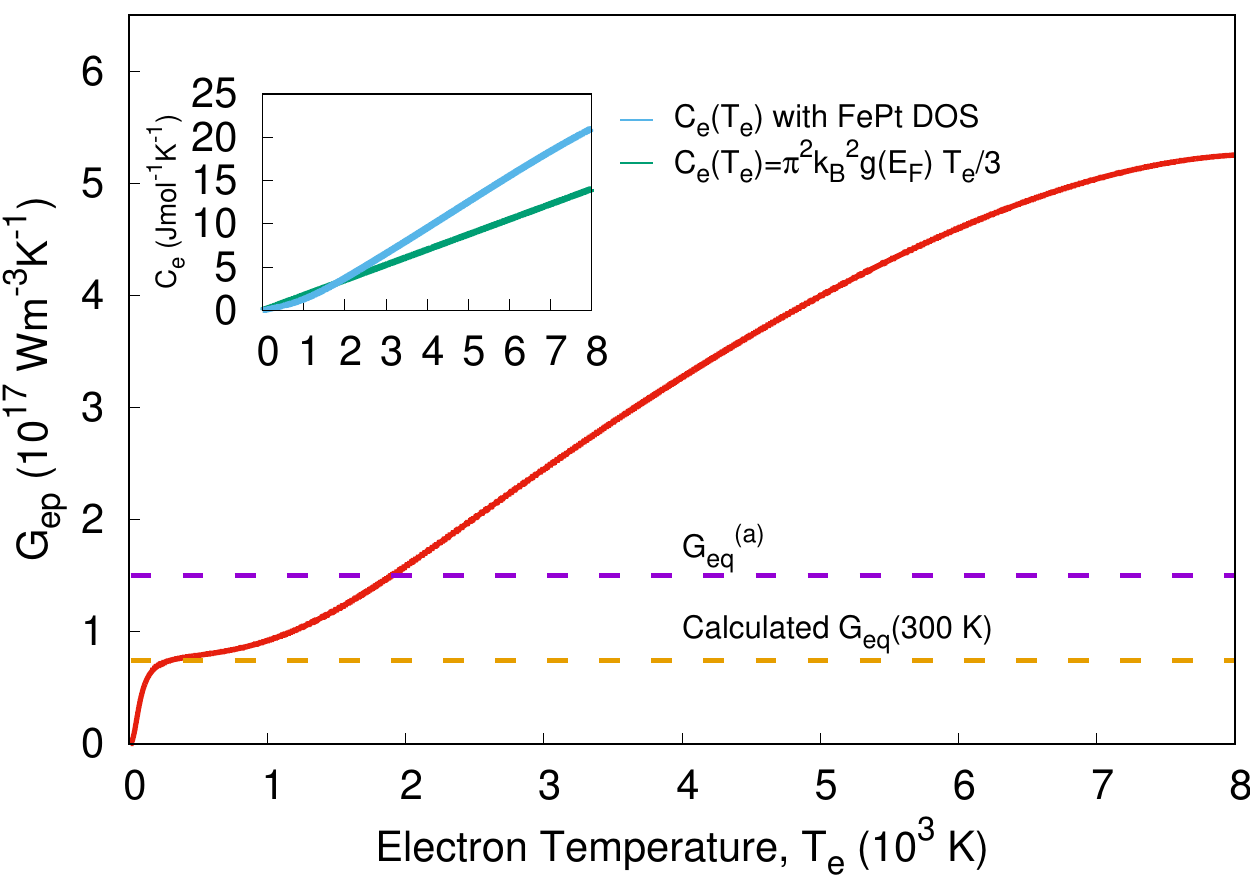}
\caption{Calculated total electron-phonon coupling as a function of electronic temperature (red line), compared with the computed total electron-phonon coupling at 300\,K (yellow line), and with a theoretical estimation  ($^{\rm a}$: purple line) \citep{Mendil2014}. The inset shows the temperature-dependent electronic specific heat.}
\label{Fig2}
\end{figure}

 To account for the scattering processes of electrons away from the Fermi surface, we have additionally included an electron temperature dependence of the mode-dependent electron-phonon coupling constant, using \cite{Lin2008}
\begin{equation}
G_Q(T_e)=G_Q\int_{-\infty}^{+\infty}d\epsilon\frac{\partial f_k}{\partial\epsilon}\frac{g(\epsilon )^2}{g^2(\epsilon_F)},
\end{equation}
with $g(\epsilon )$ being the electron density of states at  energy $\epsilon$. In Fig.\ \ref{Fig2} we illustrate the relevance of the electron temperature for $G_Q$. The calculation shows a fast growth of $\frac{1}{N_Q}\sum_QG_Q$ with $T_e$ (red line), as compared with the temperature independent value (yellow line) and an estimated value of $G_{eq}$ used recently \citep{Mendil2014} to reproduce experimental data with a 2TM (purple line). At high electronic temperatures, $G_{ep}$ is an order of magnitude larger, which will clearly influence the early dynamics of the system  after laser irradiation. 

 Also, we would like to stress that to correctly assess the dependence of the electron heat capacity on the electronic temperature, in our model we calculate it as the derivative of the total electron energy density against the electron temperature \cite{Lin2008}, 
 $C_e(T_e)=\int_{-\infty}^{+\infty}({\partial f_k}/{\partial T_e})g(\epsilon)\epsilon d\epsilon$. This is
 in contrast to the commonly used calculation of $C_e$  from the Sommerfeld expansion of the free energy, which provides a linear temperature dependence, i.e.\ $C_e(T_e)=\pi^2k_B^2g(\epsilon_F)T_e/3$. The difference between these two different descriptions is shown in the inset of Fig.\  \ref{Fig2}.



\begin{figure}[tb]
\includegraphics[width=0.99\linewidth]{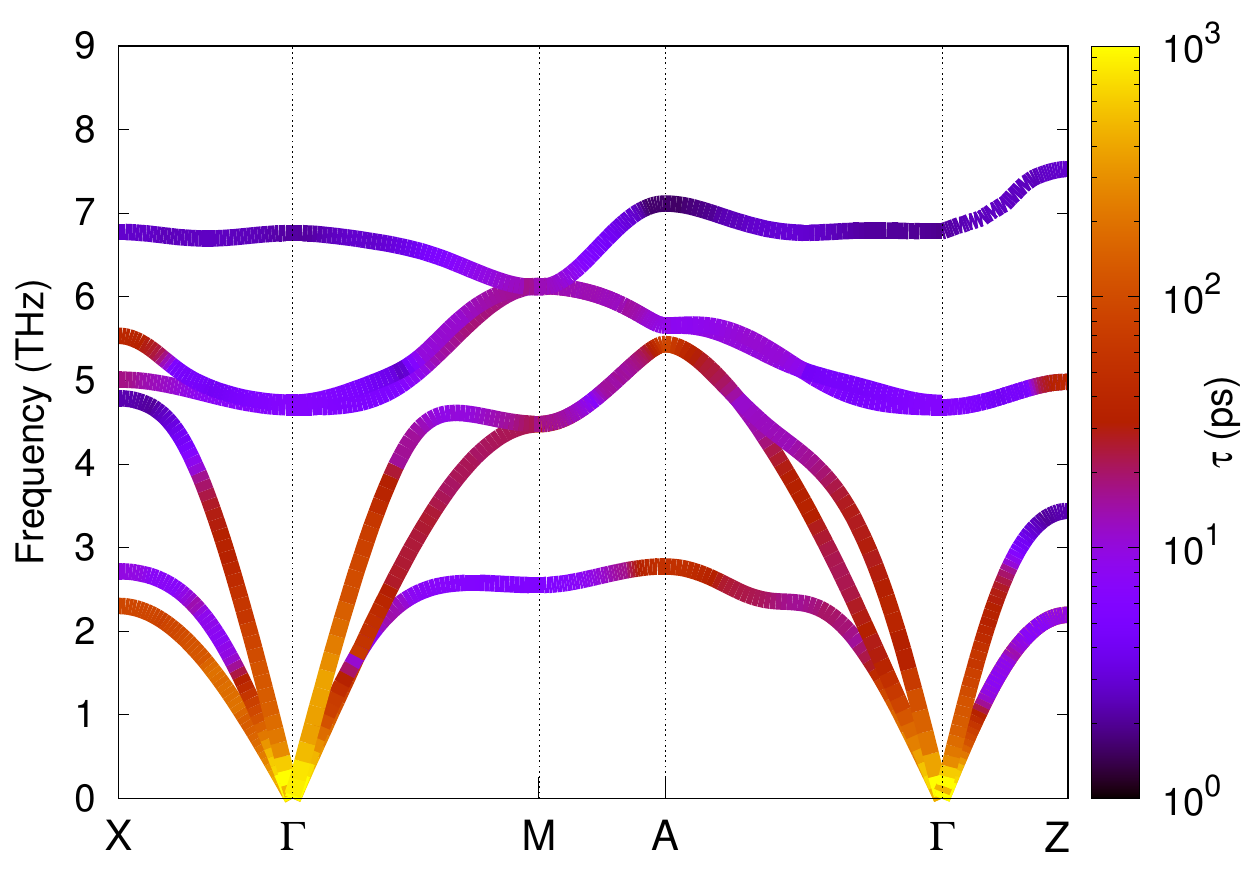}
\caption{
Calculated phonon dispersions of ferromagnetic FePt along high-symmetry lines. The symbol color is related to the phonon lifetimes at 300 K due to phonon-phonon interaction (ser color bar on the right).}
\label{Fig3}
\end{figure}

Another key quantity of our model is the explicit incorporation and calculation of the phonon anharmonicities, that enter into the model through the mode-dependent phonon linewidths $\Gamma_{Qk}$. This quantity is related to the mode-dependent phonon lifetime $\tau_Q$ due to phonon-phonon interaction, via $\tau_Q=[{2\sum_k \Gamma_{Qk}}]^{-1}$. Notably, in the 2TM it is assumed that such interactions are very strong and lead to very short phonon lifetimes, and therefore to an immediate equilibration of the lattice. In Fig.\ \ref{Fig3} we show  the  calculated $\tau_Q$  of FePt at 300\,K along high-symmetry lines in the BZ. It can be clearly seen how different phonon modes and branches have different lifetimes, as indicated in the color changing when moving along the phonon dispersions. 
The phonon lifetimes for the optical branches range from 2 to 10 ps, while the acoustic branches have phonon lifetimes larger than 10 ps, diverging at the $\Gamma$ point. These timescales are much larger than the initial ultrafast dynamics of the system. Thus, our calculations show that the assumption of an immediately thermalized lattice, as made in the 2TM, is not tenable, since the phonon-phonon lifetimes are at least one order of magnitude larger than the electron-phonon lifetimes.

The combination of the main results shown in Figs.\ \ref{Fig1} and  \ref{Fig3}, namely strong mode-dependent excitation of phonon modes via electron-phonon coupling and slow lattice thermalization through phonon-phonon interaction, suggests that the lattice remains out-of-equilibrium not only at sub-ps timescales, but even on much larger timescales. This is a striking difference with respect to the model by Waldecker \textit{et al.}\ \cite{Waldecker2016,Waldecker2017}, who predicted that phonons thermalize within a few picoseconds. The disagreement with their results is a consequence of the different  theoretical description they proposed. They use a (non-microscopically) derived phonon branch dependence to account for the different strength in the electron-phonon interaction, and obtain the phonon-phonon coupling by fitting experimental data under the assumption of equal phonon-phonon interaction strength between branches. This assumption is however not justified as we have seen in Fig.\ \ref{Fig3}, where the phonon lifetime is strongly mode dependent. Also, as we have seen in Fig.\ \ref{Fig1}, the electron-phonon coupling constants do in fact strongly vary within the same phonon branch. 

 
To provide the first \textit{ab initio} description of laser-induced nonequilibrium dynamics in a metal, and to answer the question for  how long the lattice is out-of-equilibrium, we have numerically solved our nonequilibrium rate Eqs.\ (\ref{rate-E1}) and (\ref{rate-E2}), considering $N_Q=16^3$ phonon modes (enough to achieve good numerical convergence). 

\begin{figure}[tb]
\includegraphics[width=0.99\linewidth]{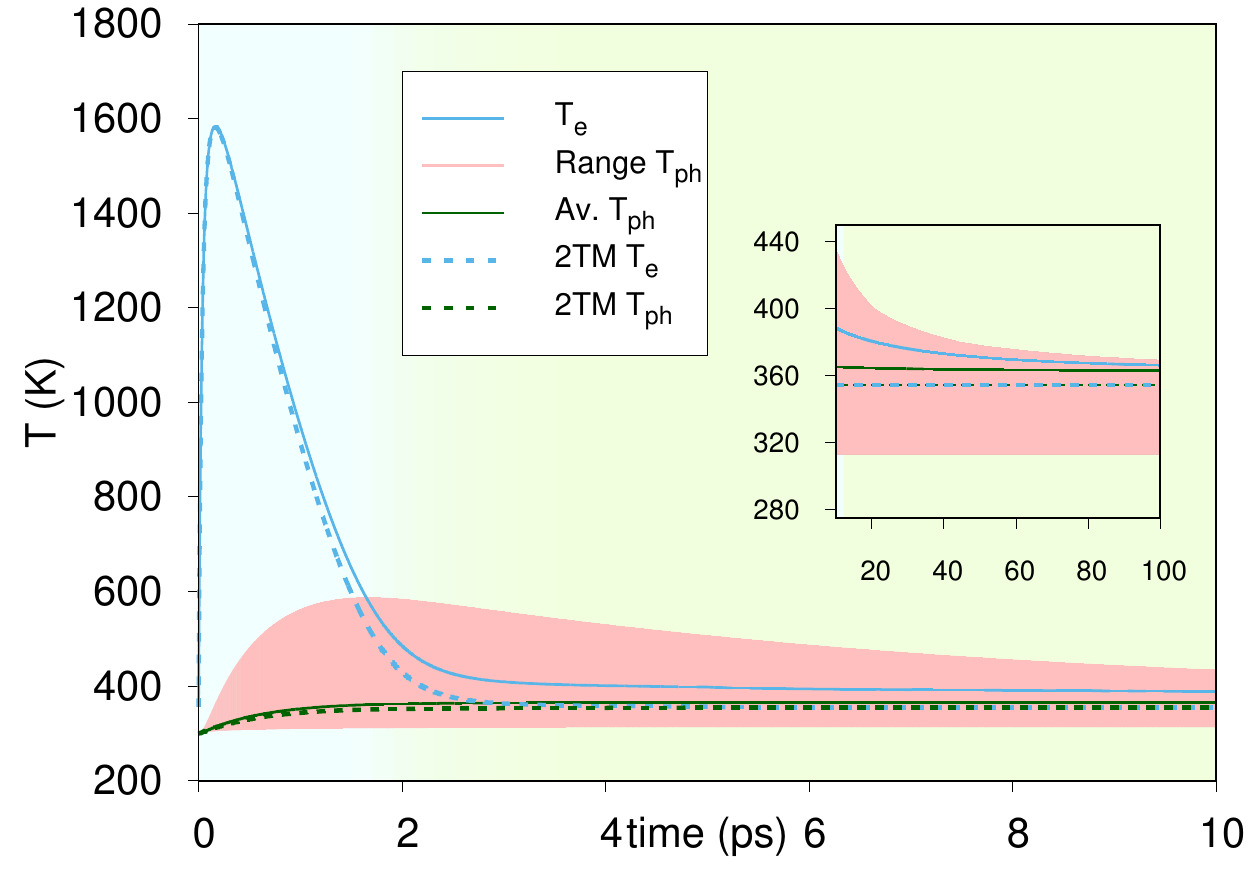}
\caption{\textit{Ab initio} calculated temporal evolution of the electronic temperature (blue line), average phonon temperature (green line) and temperature range within which all the phonon-mode temperatures are contained (red area). The inset shows the temporal evolution from 20 to 100\,ps. Dashed lines show the results of the \textit{ab initio} 2TM for FePt.}
\label{Fig4}
\end{figure}

In Fig.\ \ref{Fig4} we plot the temporal evolution of the electronic and phonon mode temperatures for times up to 100\,ps. The electronic temperature (blue line) increases rapidly, reaching up to 1575\,K in 196\,fs, followed by an exponential decrease with a decay time of about 1225\,fs (light shaded area).
The final electronic temperature reached (after 100\,ps) is 366\,K.  We also show the phonon mode-dependent range of temperatures versus time (red area). The individual evolution of each phonon mode is left out for sake of simplicity. Initially, the maximum values in the range, which stem from optical phonon modes, increase, reaching a value of about 590\,K at 1660\,fs, followed by a slow monotonic decrease. On the other hand, the minimum values, which stem from acoustic modes close to the  $\Gamma$ point, keep increasing very slowly, reaching a temperature of 313\,K at 100\,ps. Notably, the phonon mode temperatures cover a range of hundreds of Kelvins during several picoseconds, evidencing a nonequilibrium behavior. To estimate the weight of each phonon mode in the thermalization process and to avoid the singular behavior of some phonon modes, such as phonon modes close to $\Gamma$, we also show the average phonon temperature (green line), calculated as the sum of phonon mode temperatures over $N_Q$.
Although the average phonon temperature reaches an almost converged value very fast (within the first picoseconds), it is stunning to observe that this temperature still differs from the electronic temperature even at 100\,ps, which confirms a continuing energy flow between the electronic and phononic systems. 
Moreover, the phonon modes temperatures cover an interval of about 50\,K at 100\,ps. 

For comparison we have computed the electronic temperature evolution using the 2TM  with the parameters obtained recently to simulate ultrafast demagnetization in FePt \cite{Mendil2014}. The results, shown in the SM, are strikingly different. Not only does $T_e$ reach higher values (around 2500\,K) in much shorter time (about 100\,fs), also the final electronic temperature is higher, about 550\,K. Additionally, the electron-lattice dynamics is much faster, reaching a common thermal equilibrium after only 1.5\,ps. 
In contrast, our results  evidence that the lattice remains out-of-equilibrium not only during short time scales after laser excitation but also on large timescales, and provide a clear example that a complete nonthermal modeling is needed to describe the out-of-equilibrium dynamics. 



In conclusion, we have developed a nonequilibrium theory to describe the out-of-equilibrium dynamics triggered by ultrashort laser pulses. 
The model enables a fully \textit{ab initio} determination of the out-of-equilibrium dynamics. 
Our simulations for FePt unambiguously reveal that, in contrast to previous understanding,
the lattice is not in equilibrium even after 20\,ps.
As ultrafast nonequilibrium dynamics of materials is a strongly emerging research area and since our theory can provide a fully 
 parameter-free description of the ensuing dynamics,   we expect it to become a valuable tool for future modeling of ultrafast relaxation dynamics of laser-excited metals.

We thank H.\ D{\"u}rr and M.\ Bargheer for valuable discussions.
This work has been funded through the Swedish research Council (VR), the K.\ and A.\ Wallenberg Foundation (Grant No.\
2015.0060) and the R{\"o}ntgen-{\AA}ngstr{\"o}m Cluster. We also acknowledge support from the Swedish National Infrastructure for Computing (SNIC).

\bibliography{references} 

\begin{thebibliography}{41}
\bibitem [{not()}]{note1}%
  \BibitemOpen
  \href@noop {} {}\bibinfo {note} {The factor 2 stems from the spin degeneracy,
  but a more general form can be given.}\BibitemShut {Stop}%
\end{thebibliography}

\end{document}